\documentstyle[prl,aps,epsf]{revtex}

\def\Journal#1#2#3#4{{#1} {\bf #2}, #3 (#4)}

\def\NP{{\em Nucl. Phys.}}

\def\PL{{\em Phys. Lett.}}

\def\PRL{\em Phys. Rev. Lett.}
\def\PR{{\em Phys. Rev.}}
\def\PRD{{\em Phys. Rev.} D}
\def\PRC{{\em Phys. Rev.} C}
\def\ZPC{{\em Z. Phys.} C}

\def\PRep{{\em Phys. Rep.}}
\def\PPNP{{\em Prog. Part. Nucl. Phys.}}
\def\CPC{{\em Comp. Phys. Comm.}}

\def\HIP{{\em Heavy Ion Phys.}}

\newcommand{\AmS}{{\protect\the\textfont2
  A\kern-.1667em\lower.5ex\hbox{M}\kern-.125emS}}
\newcommand{\comment}[1]{}
\newcommand{\etal}{{\it et al.}}

\begin{document}
\draft
\title{Baryon distribution for high energy heavy ion collisions
        in a Parton Cascade Model}

\author{Yasushi Nara}

\address{RIKEN BNL Research Center,
   Brookhaven National Laboratory, Upton, New York, 11973, USA}

\maketitle

\begin{abstract}
The baryon distribution is studied by using
 a parton cascade model which is based on pQCD
  incorporating hard partonic scattering
  and dynamical hadronization scheme.
In order to study baryon distribution,
baryonic cluster formation is newly implemented as well as
hadronic higher resonance states from parton/beam cluster decay.
The net baryon number and charged hadron distributions
are calculated
with different $K$-factors in which parameters are fixed by elementary
$p\bar p$ data at $E_{c.m.}=200$ GeV.
It is found that baryon stopping behavior at SPS and RHIC energies
are not consequence of hard parton scattering but soft processes.
\end{abstract}


\section{INTRODUCTION}

Heavy ion experiments at BNL-AGS and CERN-SPS have been 
performed motivating by the possible creation of QCD phase transition
and vast body of systematic data such as
proton, pion strangeness particles distributions, HBT correlation, flow,
dileptons and $J/\psi$ distributions have been accumulated
 including mass dependence and their excitation functions~\cite{qm96,qm97,qm99}.
Data from forthcoming experiment at BNL-RHIC 
will be available soon.

Strong stopping of nuclei has been reported both at AGS and at SPS energies
~\cite{e866,na49B}.
It is reported that
baryon stopping power can be understood within a hadronic models
if we consider multiple scattering of nucleon
 using reasonable $pp$ energy loss~\cite{lexus}.
For example, within string based models
~\cite{venus,rqmd,hijingb,hijingbb,capella},
baryon stopping behavior at SPS energies is well explained
 by introducing diquark breaking mechanism
 in which diquark sitting at the end of the string breaks.
Diquark breaking leads to large rapidity shifts of the baryon.
Constituent quark scattering within a formation time~\cite{rqmd,urqmd} has to
be considered in order to generate Glauber type multiple collision
at initial stage of nuclear collisions in 
 microscopic transport models
 which describe full space-time evolution of particles.

Event generators based on perturbative QCD (pQCD) are proposed such as
HIJING (Heavy Ion Jet Interaction Generator)\cite{xnwang1,hijing},
VNI (Vincent Le Cucurullo Con Giginello)\cite{vni},
in order to describe ultra-relativistic heavy ion collisions
emphasizing the importance of mini-jet productions.
VNI can follow the space-time history of partons and hadrons.
The parton cascade model of VNI has been applied to study several
aspects of heavy-ion collisions even at SPS energies~\cite{vni2}.
However, original version of VNI implicitly
 assumed the baryon free region at mid-rapidity
during the formation of hadrons, because
only two parton cluster (mesonic cluster) formations are included
 in the Monte-Carlo event generator VNI~\cite{vni}.

In this work,
The baryon distribution at SPS and RHIC energy are discussed using
modified version of parton cascade simulation code VNI~\cite{vnim}.
The main features of the parton cascade model to be used here
are that implementation of baryonic cluster formation
and during the parton/beam cluster decay higher
 hadronic resonance states are allowed to produce
 in order to be able to calculate baryon distribution in heavy ion
collisions.

\comment{
The article is organized in the following way.
In section ~\ref{sec:pcm},
I summarize the main component of parton cascade model VNI
and its extensions.
In section \ref{sec:result},
I first compare some VNI results to SPS data and
discuss the baryon stopping and $\Lambda$ yield.
In section,\ref{sec:summary},
I draw conclusions.
}

\section{PARTON CASCADE MODEL}\label{sec:pcm}

First of all,
the main features of the parton cascade model
of VNI as well as the main points of the modification will be presented.
Relativistic transport equations for partons based on QCD
~\cite{geiger1} are
basic equations which are solved on the computer in parton cascade model.
The hadronization mechanism is described in terms of dynamical parton-hadron
conversion model of Ellis and Geiger~\cite{geiger2,EG1,EG2}.
The main features in the Monte Carlo procedure are summarized as follows.

1) The initial longitudinal momenta of the partons
  are sampled according to the measured nucleon
   structure function $f(x,Q_0^2)$ 
   with initial resolution scale $Q_0$.
 We take GRV94LO (Lowest order fit)~\cite{GRV}
  for the nucleon structure function.
 The primordial transverse momenta of partons are 
 generated according to the Gaussian distribution with mean value of
 $p_{\perp}=0.44$GeV.
The individual nucleons are assigned positions according to a Fermi
distribution for nuclei and the positions of partons are distributed
 around the centers of their mother nucleons with an exponential
 distribution with a mean square radius of 0.81fm.

2)With the above construction of the initial state,
the parton cascading development proceeds.
Parton scattering are simulated using closest distance approach method
in which parton-parton two-body collision will take place if
 their impact parameter becomes less than $\sqrt{\sigma/\pi}$,
where $\sigma$ represents the parton-parton scattering cross section
calculated by pQCD within a Born approximation.
Both spacelike and timelike radiation corrections
 are included within the leading
logarithmic approximation.
Elementary $2\to2$ scatterings, $1\to2$ emissions and $2\to1$ fusions
are included in the parton cascading.

3) Parton clusters are formed from secondary partons that
  have been produced by the hard interaction and parton branching.
The probability of the parton coalescence to form color-neutral
cluster $\Pi$ is defined as~\cite{EG1}
\begin{equation}
 \Pi_{ij\to C} = \left\{
     \begin{array}{lll}
        0, & L_{ij} \leq L_0,\\
        \displaystyle{
        1-\exp({L_0-L_{ij} \over L_c-L{ij}}) },&  L_0<L_{ij}\leq L_c,\\
        1, & L_{ij}> L_c,
    \end{array}\right.
\end{equation}
where $L_c=0.8$fm is the value for the confinement length scale
and $L_0=0.6$fm is introduced to account for finite transition region.
$L_{ij}$ is defined by the distance between parton $i$ and its 
nearest neighbor $j$:
\begin{equation}
  L_{ij} \equiv \min(\Delta_{i1},\cdots,\Delta_{ij},\cdots,\Delta_{in}),
\end{equation}
where $\Delta_{ij}=\sqrt{r_i^\mu r_{j\mu}}$
 is the Lorenz-invariant distance between partons.
So far, only the following two-parton coalescence
\begin{eqnarray}
g + g &\to& C_1 + C_2, g + g \to C + g, g + g \to C + g + g,\\ 
q + \bar q &\to & C_1 + C_2, q + \bar q \to  C + g, \\
q + g &\to & C + q, q + g \to  C + g + q.
\end{eqnarray}
have been considered in the VNI model.
In this work,
if diquarks are formed with the above formation probability,
 baryonic cluster formation is included as
\begin{eqnarray}
qq + q &\to& C,\\
\bar q \bar q + \bar q &\to & C, \\
q_1q_2 + \bar q_3 &\to & q_1\bar q_3 + q_2,\\
q_1q_2 + g  &\to & q_1q_2q_3 + \bar q_3.
\end{eqnarray}
Note that by introducing those cluster formation processes,
We do not introduce any new parameters into the model.

4) Beam clusters are formed from primary partons (remnant partons)
 which do not interact during the evolution even though
 they travel in the overlapping region of nuclei.
They may be considered as the coherent relics of the original hadron
wavefunctions, and should have had soft interactions.
Those underlying soft interactions are simulated by the beam cluster decay
into hadrons in VNI because additional possibility that several parton
pairs undergo soft interactions.
This may give a non-negligible contribution to the
 `underlying event structure' even at the collider energies.
The primary partons are grouped together to form a massive beam cluster
with its four-momentum given by the sum of the parton momenta and its
position given by the 3-vector mean of the partons' positions.

5) The decay probability density of each parton cluster into
final state hadrons including hadronic resonances
is chosen to be a Hagedorn density state.
The appropriate spin, flavor, and phase-space factors
are also taken into account.
In the decay of parton/beam cluster, higher hadronic resonance
states up to mass of 2GeV can be produced in our model.

To summarize, the main different points from original version are
1) baryonic cluster formation.
2) inclusion of higher hadronic resonance up to mass of 2GeV.
3) exact conservation of flavor, i.e. (baryon number, charge,etc).
4) reasonable total momentum conservation: total momentum is
conserved within 10\% at RHIC energy for central Au+Au collision.

\section{RESULTS}\label{sec:result}

\subsection{Elementary collisions}\label{subsec:result}

\begin{figure}[htb]
  \centerline{\epsfxsize=12cm \epsfbox{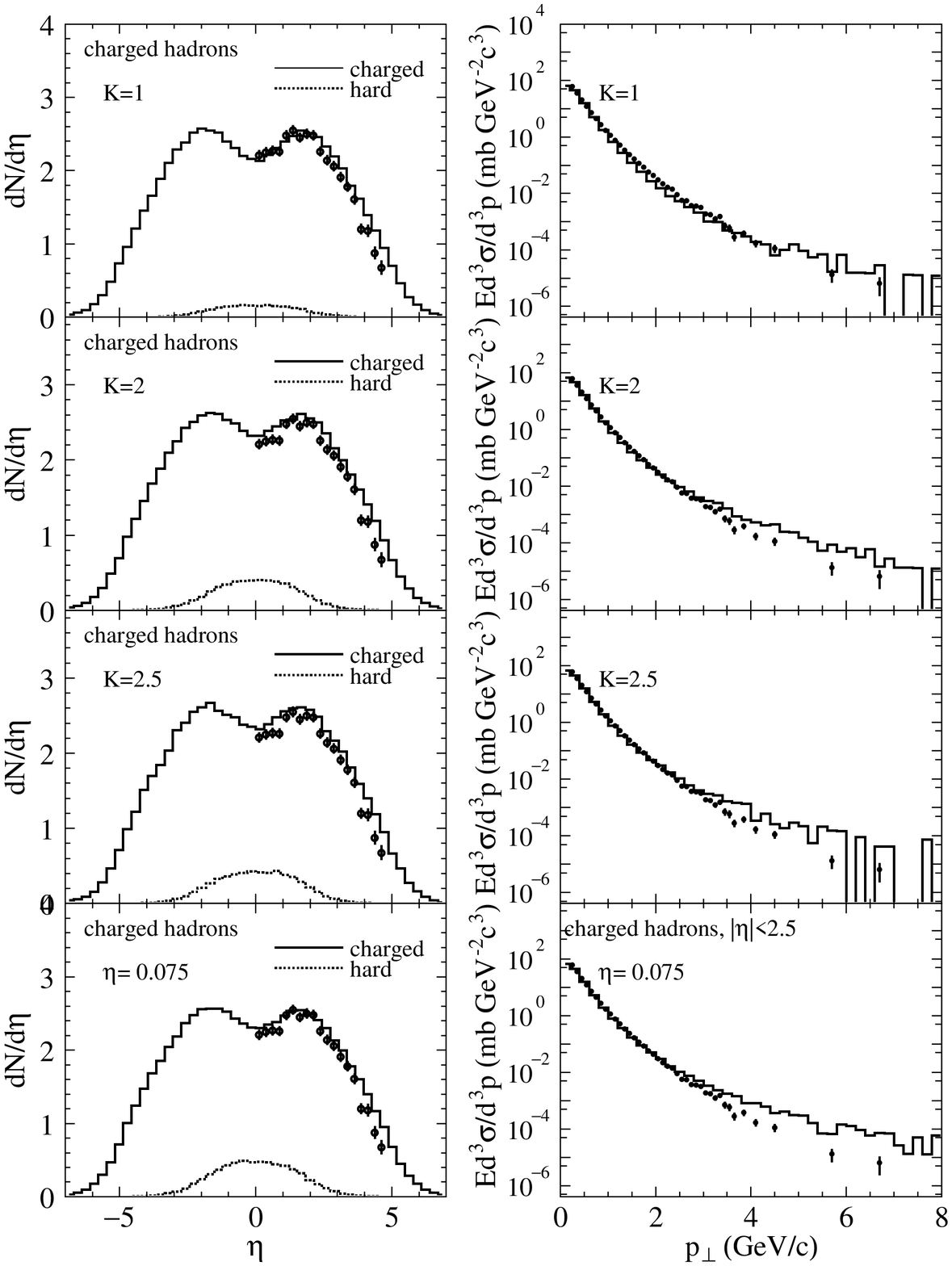}}
  \caption{
Data on charged particle ($(h^-+h^+)/2$) pseudorapidity (left)
\protect\cite{UA5e}
and the invariant cross sections $Ed^3\sigma/d^3p$ (right) \protect\cite{UA1p}
from $p\bar p$ collision at $\sqrt{s}=200$GeV
compared to parton cascade model calculations
with various parameters.
The contributions from parton cluster decay  are also plotted
by dotted lines in a left panel.
          }\label{fig:pp_exp}
\end{figure}

Since our version of parton cascade code differs from original version
of VNI, we have to check the model parameters.
First, particle spectra from $p\bar p$ collisions
at $\sqrt{s}=200$GeV
calculated by the modified version of VNI
 are studied to see the model parameter dependence.
Here we see the $K$-factor dependence as mentioned in
Ref.~\cite{bass}.
In Fig.~\ref{fig:pp_exp},
  experimental data on pseudorapidity distributions (left panel)
 and the invariant cross sections (right panel)
are compared to the calculation of the parton cascade model
with different parameters on the treatment of so-called $K$-factor.
The calculations (upper three figures) are done by adding
the constant factor to the reading-order pQCD cross sections:
$$\sigma_{pQCD}(Q^2)=K\times \sigma^{LO}(Q^2)$$
with values $K=1,2,2.5$.
While bottom figure corresponds to
the calculation changing the $Q^2$ scale in the running coupling
constant  $\alpha_s$ as
$$\sigma_{pQCD}(Q^2)=\sigma^{LO}(\alpha_s(\eta Q^2))$$
with the value $\eta=0.075$.
We also plot the contribution from parton cluster decay
in the left panel with dotted lines.
The contribution of parton cluster decay which is come from
interacted parton coalescence
 changes according to the choice of the correction scheme.
We can fit the $p\bar p$ data of pseudorapidity distributions
 with different correction schemes as seen in Fig.~\ref{fig:pp_exp}
by changing the parameter (in actual code, {\tt parv(91)})
 which controls the multiplicity from beam cluster.
We have to check the model with various elementary data
including incident energy dependence in order to fix model parameters.
Next we will present some results on nuclear collisions
with those parameters.

\subsection{Comparison with SPS data}\label{subsec:comp}

The baryon stopping problem is one of the important element
in nucleus-nucleus collisions.
Original version of VNI implicitly assumed baryon free region at midrapidity,
because baryonic parton cluster formation is not included.
Baryons only come from beam cluster, not parton cluster formation in the
original version of VNI.
We can now discuss the baryon stopping problem with our modified version
of VNI.

\begin{figure}[htb]
  \centerline{\epsfxsize=11cm \epsfbox{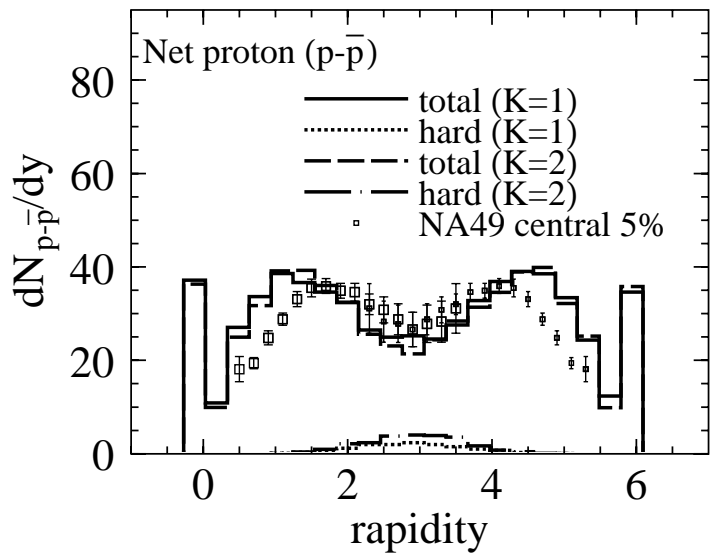}}
  \caption{Parton cascade model calculations of
        the rapidity distributions of net protons
        for Pb + Pb collision at SPS energy ($E_{lab}=158$ AGeV).
    $K=1.0$ (solid and dotted lines)
    and $K=2.0$ (dashed and dash-dotted lines)
     are used in this calculation.
Dotted and dash-dotted lines corresponds to the contribution from
 parton cluster decay respectively.
          }\label{fig:pbpb_rap}
\end{figure}

We have calculated the net proton distribution at SPS energy
to show the reliability of the modeling of beam cluster formation
 in the parton cascade model.
Fig.~\ref{fig:pbpb_rap} compares the parton cascade calculation
for Pb+Pb collision at the laboratory energy of $E_{lab}=158$ AGeV
with the $K$-factor 1.0 (original version uses $\eta=0.035$)
 of net protons with the data~\cite{na49B}.
It is seen that contribution from parton cluster is neglibigly
small, thus baryon stopping behavior is fully explained by
soft physics (in this case, beam cluster decay) when
we chose the $K$-factor 1.0 at SPS energies.
It should be noted that
there is no microscopic dynamics
in the modeling of the beam cluster formation
 in the parton cascade model, but it is a simple fit to the data
of $pp$ collisions.

\subsection{Predictions for RHIC}\label{subsec:comprhic}

\begin{figure}[htb]
  \centerline{\epsfxsize=12cm \epsfbox{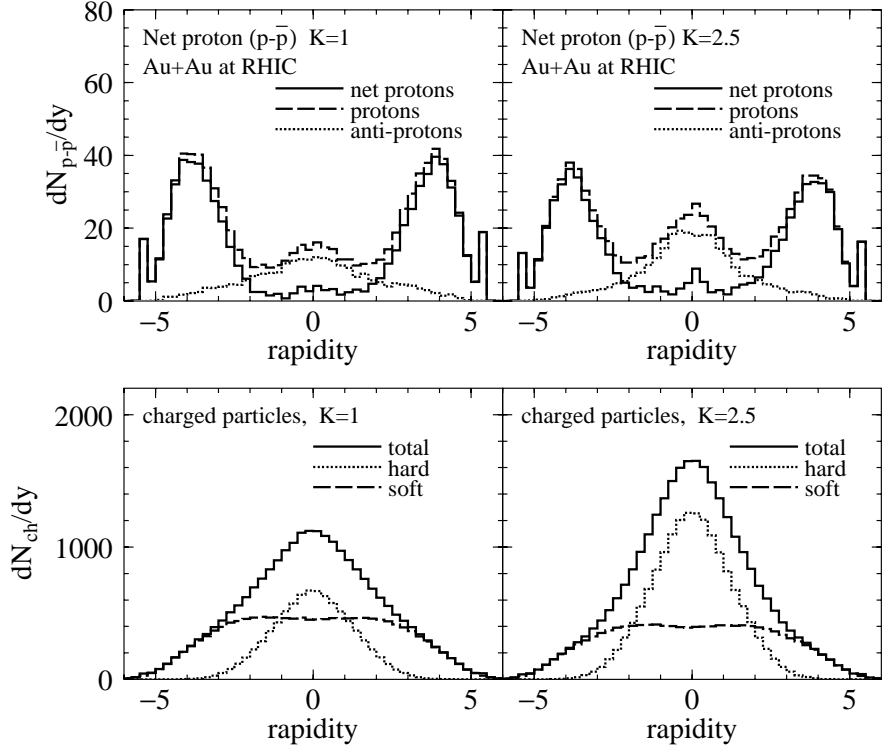}}
  \caption{Parton cascade model calculations of
        the rapidity distributions of net protons, protons,
and anti-protons (upper)
        and charged particles (lower)
        for Au + Au collision at $E_{c.m.}=200$ AGeV
  for head on collisions.
          }\label{fig:rap}
\end{figure}

The $K$-factor dependence of both net proton
and charged particle rapidity distribution are studied
 in Fig.~\ref{fig:rap}
In terms of net proton distribution, there is no strong $K$-factor
dependence. 
We can see that parton cluster formation and its decay
predict almost baryon free at mid-rapidity region regardless
of the choice of $K$-factor, though there are lots of protons
and antiprotons at mid-rapidity.
We conclude that hard parton scattering plays no rule for
the baryon stopping within a parton cascade model.
However, note that string based model like HIJING/B~\cite{hijingb,hijingbb}
predicts proton rapidity density of 10
and UrQMD predicts~\cite{bleicher} 12.5 at mid-rapidity.
However, as pointed out in Ref.~\cite{bass},
charged hadron multiplicity is strongly depend on how to chose
the leading order correction scheme.

\begin{figure}[htb]
  \centerline{\epsfxsize=16cm \epsfbox{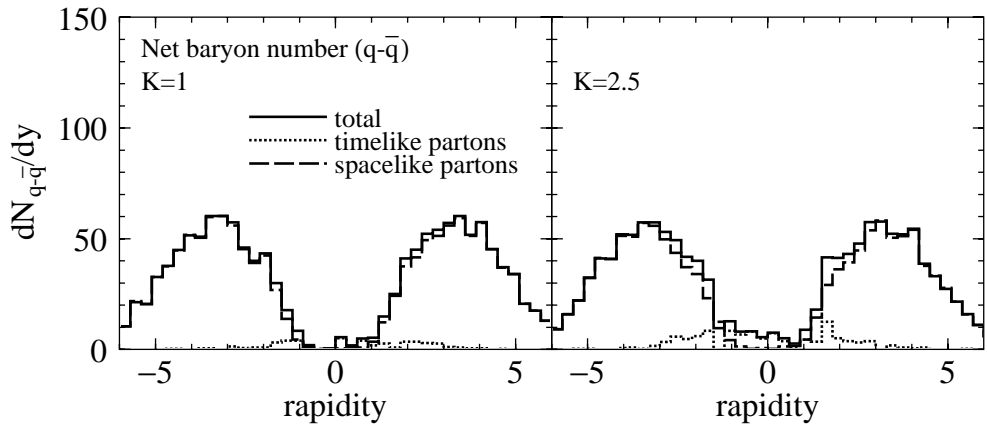}}
  \caption{
 Rapidity distributions of net baryon number ($q-\bar q$)
 obtained from parton cascade model before hadronization
 with $K=1$ (left) and $K=2.5$ (right) 
 for Au + Au collision at $E_{c.m.}=200$ AGeV for head on collisions.
          }\label{fig:rap_netq}
\end{figure}

Fig.~\ref{fig:rap_netq} displays the net baryon number distributions
as a function of rapidity obtained by parton distribution
from parton cascade before hadronization
with the $K$-factor of 1 (left) and 2.5 (right).
Net baryon number of time-like partons are distributed
around mid rapidity region but its contribution are small
as consistent with the net proton distribution in Fig.~\ref{fig:rap}.

\section{SUMMARY}\label{sec:summary}

In summary,
first, we have checked that different treatments for the
inclusion  of higher-order pQCD corrections in parton cascade
model can fit the elementary $p\bar p$ collisions.
We have to check other elementary processes to fix
the model parameters.
We show the net proton rapidity distribution at SPS energies
to demonstrate that the beam cluster
 treats underlying soft physics in the parton cascade model
reasonably well for nucleus nucleus collisions. 
Then,
we have calculated the net proton rapidity distribution at RHIC
energy as well as charged particle distributions
using modified version of parton cascade code VNI in which
we newly introduced  baryonic parton cluster formation
and higher hadronic resonance states from decay of parton
and beam cluster.
Within a framework of perturbative parton cascading and
dynamical hadronization scheme,
we predict almost baryon free plasma at RHIC energy.
The charged particle rapidity distributions are also
studied with the parameter set which are fitted by $p\bar p$
collisions.  Strong $K$-factor dependence on the
hadron multiplicity is seen as
previously being found by Ref.~\cite{bass}.
we can not fix the $K$-factor from
 only rapidity and transverse momentum distributions
for $p \bar p$ collisions.

In this work, we consider only two or three parton coalescence,
but in dense parton matter produced in heavy ion collisions,
this assumption might be broken down.
Inverse processes like hadron conversion to parton such as
 $C \to q \bar q$ are also ignored which might become
 important at higher colliding energies.

\section*{ACKNOWLEDGMENTS}
This work should have been collaborated with Klaus Geiger
if he had not had perished in the air crash.
I would like to thank Dr. S. A. Bass and  Prof. R. S. Longacre
for careful reading of this paper and useful comments.
I am indebted to S. Ohta for encouragements and useful comments.

\end{document}